\documentclass[12pt,a4paper]{article}
\usepackage{graphicx} 
\usepackage{url}      
\newcommand{\doi}[1]{\url{http://dx.doi.org/#1}}

\usepackage{subfigure}
\usepackage{multirow}
\usepackage{amsmath,amssymb,latexsym,amsfonts}

\newcommand{\ignore}[1]{ }
\usepackage{color} 

\newtheorem{df}{Definition}[section]
\newtheorem{corollary}[df]{Corollary}
\newtheorem{remark}[df]{Remark}

\let\ro=\varrho

\def\R{\mathbb R}

\title{Comparison of numerical and analytical approximations of the early exercise boundary of the American put option}

\author{
Martin~Lauko\thanks{Dept.\ Applied Mathematics \& Statistics, Comenius University, 842 48  Bratislava, Slovakia.
{\tt sevcovic@fmph.uniba.sk}}
\and 
Daniel~\v{S}ev\v{c}ovi\v{c}\footnotemark[1] 
}

\date{}

\begin{document}

\maketitle

\begin{abstract}
In this paper we present qualitative and quantitative comparison of various analytical and numerical approximation methods for calculating a position of  the early exercise boundary of the American put option paying zero dividends. First we analyze their asymptotic behavior close to expiration. In the second part of the paper, we introduce a new numerical scheme for computing the entire early exercise boundary. The local iterative numerical scheme is based on a solution to a nonlinear integral equation. We compare numerical results obtained by the new method to those of the projected successive over relaxation method and the analytical approximation formula recently derived by Zhu. 


\smallskip
\noindent {\bf AMS-MOS classification:} 35K15, 35K55, 90A09, 91B28

\end{abstract}




\section{Introduction}
\label{sec:intro}

The analysis of the early exercise boundary and the optimal stopping time for American put options on assets paying zero dividends has attracted a lot of attention from both theoretical as well as practical point of view. An American put option is a financial contract between the writer and the holder of the option. It gives the holder the right, but not the obligation, to sell the underlying asset at the prescribed strike price any time before expiration. Under the standard assumptions made on the underlying stock process and completeness of the financial market (c.f. \cite{H,Kw}) the American put option can be priced using the Black-Scholes equation (c.f. \cite{BS}) on a time dependent domain of the underlying asset price. More precisely, the early exercise boundary problem for the American put option can be formulated as follows: find a solution $V=V(S,t)$ and the early exercise boundary position $S_f=S_f(t)$ satisfying 
\begin{eqnarray}
&&\frac{\partial V}{\partial t} +  r S\frac{\partial V}{\partial S} + {\sigma^2\over 2} S^2 \frac{\partial^2 V}{\partial S^2} - r V =0\,,
\qquad 0<t<T,\ S_f(t) < S <\infty \,,
\nonumber
\\
&&V(+\infty ,t)=0,\ V(S_f(t), t)= E - S_f(t)\,, \ \frac{\partial V}{\partial S}(S_f(t),t)=-1\,,
\\
&&V(S,T)=(E-S)^+\,.
\nonumber
\label{amer-put}
\end{eqnarray}
The solution $V(S,t)$ is defined on a time-dependent domain $S\in(S_f(t), \infty )$, where $t\in(0,T)$ (cf. Kwok \cite{Kw}). 
Here $S>0$ stands for the underlying stock price, $E>0$ is the exercise (strike) price, $r>0$ is the risk-free rate, $\sigma>0$ is the volatility of the underlying stock process and $T$ denotes the time of maturity. In what follows, we denote by $\tau= T-t$ the time to maturity. The function $[0,T]\ni t \mapsto S_f(t)\in \R$ represents the early exercise boundary position. The above mathematical formulation of the problem of pricing the American put option by means of a solution to the free boundary problem is a basis for development of various integral equations for describing the early exercise boundary position $S_f(t)$. The analytical approximation formulae are often based on approximation of a solution to such an integral equation. Notice that there are also other numerical methods for approaching the free boundary problem (\ref{amer-put}) like e.g. front-fixing and transformation methods. We refer the reader to papers by Kwok and Wu \cite{KW}, \v{S}ev\v{c}ovi\v{c} \cite{Se1,Se2}, Ankudinova and Ehrhardt \cite{AE1} and references therein. 
Following Kwok \cite{Kw}, a solution $V=V(S,t)$ to the problem of pricing the American put option fulfills the following variational inequality: 
\begin{eqnarray}
\label{var-amer-put}
&&\frac{\partial V}{\partial t} +  r S\frac{\partial V}{\partial S} + {\sigma^2\over 2} S^2 \frac{\partial^2 V}{\partial S^2} - r V  \le 0\,,\quad V(S,t) \ge V(S,T)\,,
\nonumber
\\
&& \left(\frac{\partial V}{\partial t} +  r S\frac{\partial V}{\partial S} + {\sigma^2\over 2} S^2 \frac{\partial^2 V}{\partial S^2} - r V \right) \Big(V(S,t)-V(S,T)\Big) = 0, 
\nonumber
\\
&& \hbox{for all}\ \ 0<t<T,\ 0< S< \infty, 
\\
&& V(0, t)= E, \quad V(+\infty,t)=0,\quad \hbox{for} \ \ 0<t<T\,,
\nonumber
\\
&&V(S,T)=(E-S)^+\,,\quad \hbox{for} \ \ 0< S< \infty.
\nonumber
\end{eqnarray}
The formulation of the problem of pricing American put option as a variational inequality is often used when we need to compute not only the free boundary position  $S_f(t)$ but also the entire solution $V(S,t)$. The above variational inequality can be effectively solved by means of the so-called projected successive over relaxation (PSOR) method by Elliot and Ockendon \cite{EO}.

In the last decades, many different, but equivalent, integral equations for pricing the American put option  have been derived by Barone-Adesi and Whaley \cite{BW}, Bunch and Johnson \cite{BJ}, Carr, Jarrow and Mynemi \cite{CJM}, MacMillan \cite{Mac} and others. The asymptotic analysis often leads to an approximate expression  of the free boundary close to expiry. Since the closed form analytical formula for the early exercise boundary position is not known,  many  authors (see e.g. Geske, Johnson and Roll \cite{GJ,GR}, Johnson \cite{J}, Karatzas \cite{K1}, Evans, Kuske and Keller \cite{KK,EKK}, Mynemi \cite{M} and recent papers by Alobaidi \emph{et al.} \cite{A,MA}, Stamicar \emph{et al.} \cite{SSC}, the survey paper by Chadam \cite{Ch} and other references therein) investigated various approximation models and derived different approximate expressions for valuing American call and put options. We also refer to the books by Kwok \cite{Kw} and Wilmott \emph{et al.} \cite{WDH} for a survey of classical theoretical and computational results in the field of pricing the American put option.

In this paper, we focus on comparison of the valuation formulae due to Evans, Kuske and Keller \cite{KK,EKK}, Stamicar, \v{S}ev\v{c}ovi\v{c} and Chadam \cite{SSC} and the recent analytic approximation formula by Zhu  \cite{Zhu2006} (see also \cite{Zhu2007,Zhu2008}). Our main goal is to present qualitative and quantitative comparison of the above mentioned analytical and numerical approximation methods for calculating the early exercise boundary position. In the first part of the paper, we analyze and compare  asymptotic behavior of the early exercise boundary close to expiry for analytical approximations developed by  Stamicar, \v{S}ev\v{c}ovi\v{c} and Chadam \cite{SSC}, Evans, Kuske, Keller  \cite{KK,EKK} and Zhu \cite{Zhu2006,Zhu2007,Zhu2008}. 
We show that the approximation formulae due to Evans, Kuske and Keller \cite{KK,EKK} and  Stamicar, \v{S}ev\v{c}ovi\v{c} and Chadam \cite{SSC} have the same asymptotic behavior of $S_f(t)$ as $t\to T$. We also show that the analytic approximation formula due to Zhu has an asymptotic behavior differing from the previous ones by a logarithmic factor.
In the second part  we propose a new numerical scheme for computation of the entire function $S_f(T), t\in[0,T]$, based on a solution to the nonlinear integral equation from \cite{SSC}. We compare numerical results obtained by the new numerical method to those of the projected successive over relaxation method by Elliot and Ockendon \cite{EO} for solving the variational inequality (\ref{var-amer-put}) and the analytical approximation formula recently developed by Zhu \emph{et al.} in \cite{Zhu2006,Zhu2007,Zhu2008}.

\section{Analytical approximate valuation formulae}

In this section we present a survey of analytical, implicit integral and numerical approximation schemes for computing the early exercise boundary for the American put option. First we focus on the recent result due to Zhu who in \cite{Zhu2006} derived a closed analytic approximation formula for the early exercise boundary position $S_f(t)=\varrho(T-t)$. We also derive the asymptotic behavior of $S_f(t)$ for $t\to T$.
Next we concentrate on implicit representation formulae for $\varrho(\tau)$ expressed in the form of a single nonlinear integral equation for the function $\varrho$. We recall implicit integral equation derived by Stamicar, \v{S}ev\v{c}ovi\v{c} and Chadam in \cite{SSC}. We again derive the asymptotic behavior of the early exercise boundary position as $t\to T$. In the last subsection we present another approximations derived by Evans, Kuske and Keller \cite{KK,EKK}.

\subsection{Analytical approximation valuation formula by Zhu}

In this section we recall a recent interesting result due to Zhu. 
In \cite{Zhu2006} Zhu derived a new analytical approximation formula of the early exercise boundary by application of the Laplace and inverse Laplace integral transforms to a dimensionless form of the governing parabolic PDE and successfully obtained a closed analytic approximation formula for the early exercise boundary position as a sum of a perpetual option and integral that valuates early exercise boundary position. The resulting formula for the early exercise boundary $S_f(t) = \varrho(T-t)$ reads as follows:
\begin{equation}
\varrho^{Zhu}(\tau) = \frac{\gamma E}{1+\gamma} + \frac{2 E}{\pi} \int_{0}^{\infty} \frac{\zeta e^{-\tau\frac{\sigma^2}{2}(a^2+\zeta^2)}}{a^2+\zeta^2} e^{-f_1^*(\zeta)}\sin(f_2^*(\zeta)) d\zeta ,
\label{eq:zhu}
\end{equation}
where $ \gamma = \frac{2r}{\sigma^2}, \ \  a = \frac{1+\gamma}{2}, \ \ b = \frac{1-\gamma}{2}$, and 
\begin{eqnarray}
\label{eq:f12}
f_1^*(\zeta) &=& \frac{1}{b^2+\zeta^2} \left[ b \ln \left(\frac{1}{\gamma}\sqrt{a^2+\zeta^2}\right) + \zeta \arctan(\zeta/a)  \right],
\nonumber \\
f_2^*(\zeta) &=& \frac{1}{b^2+\zeta^2} \left[ \zeta \ln \left( \frac{1}{\gamma}\sqrt{a^2+\zeta^2}\right) 
- b \arctan(\zeta/a)  \right].
\\
\nonumber
\end{eqnarray}
Notice that the first summand in (\ref{eq:zhu}) represents the constant value of a perpetual put option i.e. the limit $\lim_{\tau\to\infty} \varrho(\tau)=\gamma E/(1+\gamma)$. 

\subsubsection*{Early exercise boundary asymptotic close to expiry}

\begin{figure}
\begin{center}
\includegraphics[width=6.5cm]{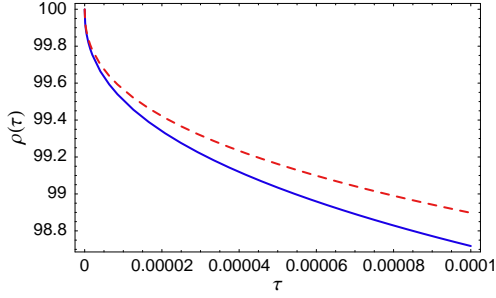}
\end{center}
\caption{Comparison of the analytic solution $\varrho^{Zhu}$ (solid curve) and its asymptotic approximation (\ref{eq:asymptotic-Zhu}) (dashed curve).}
\label{fig:asymptotic-Zhu}
\end{figure}

Next we examine the asymptotic behavior of the function $\varrho^{Zhu}(\tau)$ for $\tau\to 0$. 
Notice that we have $\varrho(0)=S_f(T)=E$ (c.f. Kwok \cite{Kw}). We shall prove that  
\[
\lim_{\tau\to 0^+} 
\frac{ E -\varrho^{Zhu}(\tau)}{\sqrt{\tau}(-\ln\tau)} = \frac{1}{\sqrt{2\pi}} E \sigma.
\]
Indeed, if we introduce the change of variables: $s= \tau\frac{\sigma^2}{2}(a^2+\zeta^2)$ we obtain
\[
\frac{E -\varrho^{Zhu}(\tau)}{\sqrt{\tau}(-\ln\tau)} =
\frac{2 E}{\pi} \int_{ \tau\frac{\sigma^2}{2}a^2}^{\infty} \frac{1-e^{-s}}{2s} e^{-f_1^*} \frac{\sin(f_2^*)}{\sqrt{\tau}(-\ln\tau)}ds, \quad\hbox{for any}\ \tau\in (0,T],
\]
where $f_i^*= f_i^*((\frac{2s}{\tau\sigma^2} -a^2)^{\frac{1}{2}}), i=1,2$. It is easy to verify that 
\begin{eqnarray*}
&&\lim_{\tau\to0^+} f_1^* = 0,\ \ \lim_{\tau\to0} f_2^* = 0,
\\
&&\lim_{\tau\to0^+} \frac{\sin(f_2^*)}{\sqrt{\tau}(-\ln\tau)} =\lim_{\tau\to0} \frac{f_2^*}{\sqrt{\tau}(-\ln\tau)} = \frac{\sigma}{2\sqrt{2s}}\,,
\end{eqnarray*}
for any $s>0$. Using the Lebesgue dominated convergence theorem we finally obtain 
\[
\lim_{\tau\to 0^+} 
\frac{E -\varrho^{Zhu}(\tau)}{\sqrt{\tau}(-\ln\tau)} = 
\frac{E\sigma }{\pi} \int_{0}^{\infty} \frac{1-e^{-s}}{(2s)^{\frac{3}{2}}} ds
=\frac{1}{\sqrt{2\pi}} E \sigma\,,
\]
as claimed. As a consequence of the previous result we can conclude the following asymptotic approximation of the formula by Zhu:
\begin{equation}
\varrho^{Zhu}(\tau) \approx E \left(1 - \frac{\sigma}{\sqrt{2\pi}} \sqrt{\tau}(-\ln\tau) \right) \ \hbox{for} \ 0<\tau\ll 1, 
\label{eq:asymptotic-Zhu}
\end{equation}
i.e. $\varrho^{Zhu}(\tau) = E \left(1 - \frac{\sigma}{\sqrt{2\pi}} \sqrt{\tau}(-\ln\tau) \right)  + o(\sqrt{\tau}(-\ln\tau))$ as $\tau\to0^+$.
In Fig.~\ref{fig:asymptotic-Zhu} we present a comparison of the analytic solution $\varrho^{Zhu}(\tau)$ and its asymptotic approximation  (\ref{eq:asymptotic-Zhu}) for $\tau\in[0, T]$ and $E=100,\sigma=0.3, r=0.1, T=10^{-4}$.

\subsubsection*{Convexity of the early exercise boundary obtained from Zhu's formula}

One of the important features of the early exercise boundary for the American put option is the convexity of the function $\varrho(\tau) = S_f(T-\tau)$ for $\tau\in (0,T]$. The analytic proof of the convexity of $\varrho$ has been presented just recently by Chadam \emph{et al.} in \cite{CCJZ}. We also recall that the early exercise boundary is log-concave as a function of log of the underlying asset price (cf.  Ekstr\"om and Tysk \cite{E,ET}). 

A relatively simple proof of the convexity of $\varrho=\varrho^{Zhu}$ follows directly from the analytic valuation formula (\ref{eq:zhu}). Indeed, for any $0<\tau\le T$, we have the following expression for the second derivative  of the function $\varrho^{Zhu}(\tau)$: 
\[
\frac{d^2}{d\tau^2}\varrho^{Zhu}(\tau) = \frac{2 E \sigma^4}{4 \pi} \int_{0}^{\infty} 
(a^2+\zeta^2)\zeta e^{-\tau\frac{\sigma^2}{2}(a^2+\zeta^2)} e^{-f_1^*(\zeta)}\sin(f_2^*(\zeta)) d\zeta.
\]
In what follows, we shall prove $f_2^*(\zeta)\equiv f_2^*(\zeta; \gamma)\in[0,\pi]$ provided that $\gamma\ge\gamma_0$ where $\gamma_0>0$  is a constant.

\begin{equation}
\gamma_0= \min(\gamma>0\ | \ \max_{\zeta>0}f_2^*(\zeta,\gamma) \le \pi )
\end{equation}

\begin{figure}
\begin{center}
\includegraphics[width=5.5cm]{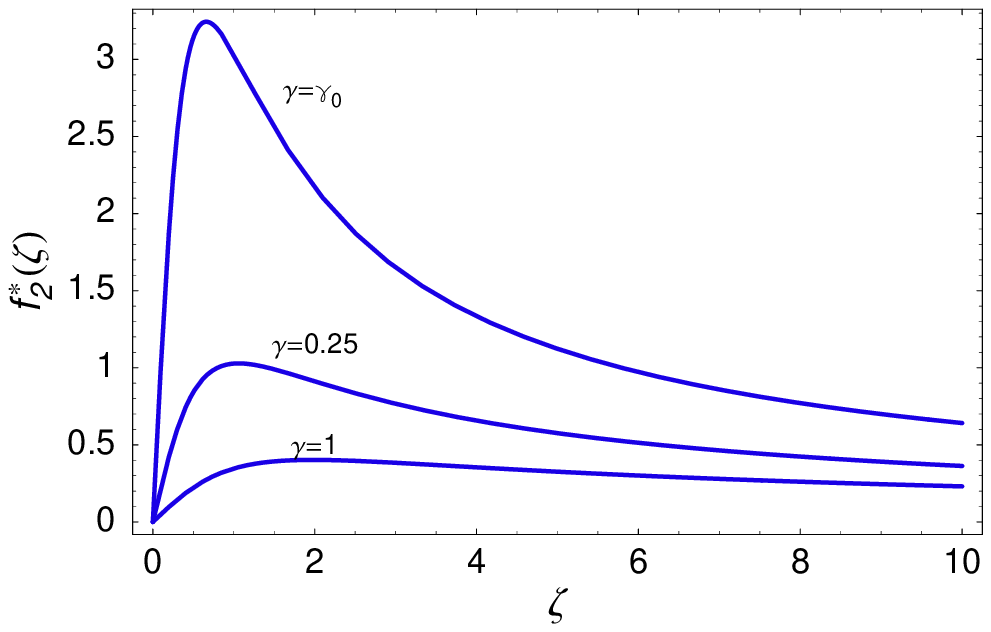}
\includegraphics[width=5.5cm]{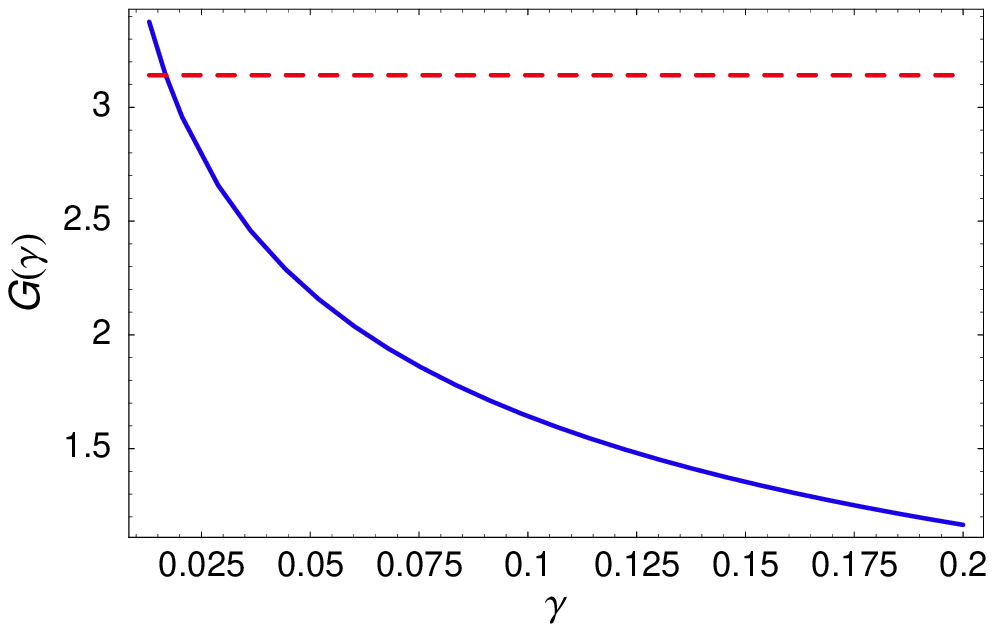}
\end{center}
\caption{A graph of the function $f_2^*=f_2^*(\zeta; \gamma)$ for various values of the parameter $\gamma$ (left). A graph of the function $G(\gamma)=\max_{\zeta>0}f_2^*(\zeta; \gamma)$ (right).}
\label{fig:functionf2}
\end{figure}

The numerical value of $\gamma_0$ can be estimated as $\gamma_0\approx 0.0167821$. 
\begin{corollary}
If $\frac{2r}{\sigma^2}=\gamma \ge \gamma_0$ where $\gamma_0\approx 0.0167821$ then $f_2^*=f_s^*(\zeta,\gamma)\in[0,\pi]$ for any $\zeta>0$. As a consequence, we have $\frac{d^2}{d\tau^2}\varrho^{Zhu}(\tau)>0$, i.e. the function $\varrho^{Zhu}(\tau)$ as well the early exercise boundary $S_f(t)$ for the  American put option are convex functions.
\end{corollary}

\begin{remark}
Notice that the condition $\frac{2r}{\sigma^2}=\gamma \ge \gamma_0$ is fulfilled for typical market-based  choices of the model parameters $r$ and $\sigma$. For example, if $r=0.01$ (i.e. $r=1\%$ p.a.) then $\frac{2r}{\sigma^2}\ge \gamma_0$ provided that the condition $\sigma^2 <1.19$ (i.e. $\sigma^2 \le 119\%$ p.a.) is satisfied.
In Fig.\ref{fig:functionf2} we present graphs of the function  $\zeta\mapsto f_2^*(\zeta; \gamma)$ for various values of the parameter $\gamma$, including the critical value $\gamma=\gamma_0\approx 0.0167821$ for which the function $G(\gamma)=\max_{\zeta>0}f_2^*(\zeta; \gamma)$ attains the critical value $G(\gamma_0)=\pi$. 
\end{remark}

\subsection{Approximation formula due to Stamicar, \v{S}ev\v{c}ovi\v{c} and Chadam}

In \cite{SSC} Stamicar, \v{S}ev\v{c}ovi\v{c} and Chadam  derived a single nonlinear integral equation for the  early exercise boundary position. Based on this integral equation the authors derived improved  analytical approximation of free boundary near the expiry. Asymptotic behavior and justification of the early exercise behavior close to expiry have been recently analyzed  by Chadam \emph{et al.} in \cite{CCJZ,CC}. They proved that the right asymptotic expansion can be obtained from the nonlinear integral equation developed by Stamicar, \v{S}ev\v{c}ovi\v{c} and Chadam in \cite{SSC}. We briefly recall key steps of derivation of the nonlinear integral equation for the early exercise boundary position $\varrho(\tau) = S_f(T-\tau)$ for the free boundary problem (\ref{amer-put}). Let us introduce the following change of variables 
$x=\ln\left(S/\ro(\tau)\right)$ where $\tau=T-t, \ro(\tau)=S_f(T-\tau)$.
Similarly as in the case of a call option (see \cite{Se1}) we define a synthetised portfolio $\Pi$ for the put option $\Pi(x,\tau) = V(S,t) - S \frac{\partial V}{\partial S}(S,t)$. Then it is easy to verify that $\Pi$ is a solution to the following parabolic equation:
\begin{eqnarray}
&&\frac{\partial \Pi}{\partial \tau}  
- a(\tau )\frac{\partial \Pi}{\partial x}
- \frac{\sigma^2}{2}\frac{\partial^2 \Pi}{\partial x^2}
+ r \Pi  = 0,\quad x>0,\tau\in(0,T),    \nonumber \\
&&\Pi (0,\tau ) = E, \quad  \Pi (\infty ,\tau ) = 0, \quad
\Pi (x,0) = 0, \quad x>0, \tau\in(0,T),
\label{bc1} \\
&&\frac{\sigma^2}{2} \frac{\partial \Pi}{\partial x}(0,\tau ) = - r E,\quad \hbox{for}\ \tau\in(0,T),\nonumber
\end{eqnarray}
where $a(\tau ) =\frac{\dot{\varrho}(\tau )}{\varrho (\tau )}+r - \frac{\sigma^2}{2}$ (see Stamicar \emph{et al.} \cite{SSC}, or \v{S}ev\v{c}ovi\v{c} \cite{Se1,Se2}). 
Applying the Fourier transform one can find the Fourier image of the function $\Pi$ in terms of the free boundary position $\varrho$. The resulting equation for the free boundary position reads as  
$\frac{\sigma^2}{2} \frac{\partial \Pi}{\partial x}(0,\tau ) = - r E$, from which the weakly singular integral equation for the function $\varrho$ can be found by using the inverse Fourier transform  (see \cite{SSC} for details). More precisely, the function $\varrho(\tau)$ fulfills the equation:
\begin{equation}
\varrho (\tau )=Ee^{-(r-\frac{\sigma^2}{2})\tau  + \sigma \sqrt{2 \tau }\eta (\tau )},
\label{eq:SSC}
\end{equation}
where the auxiliary function $\eta (\tau )$ is a solution to the following nonlinear integral equation
\begin{equation}
\eta (\tau ) =-\sqrt{-\ln \left[ \frac{r\sqrt{2 \pi \tau}}{\sigma} e^{ r\tau }\left(
1-\frac{F_\eta(\tau )}{\sqrt{\pi }}\right) \right] },\qquad \hbox{for}\ \tau\in[0,T].
\label{eq:SSCintegralequation}
\end{equation}
Here the function $F_\eta$ depends on $\eta$ via the expression
\begin{eqnarray}
F_\eta(\tau ) &=&2\int_0^{\pi /2}e^{-r\tau \cos ^2\theta 
- g^2_\eta(\tau ,\theta)}
\left( 
\frac{\sigma\sqrt{\tau }}{\sqrt{2}}\sin \theta +g_\eta(\tau ,\theta )\tan \theta 
\right)
\,d\theta ,  \label{F}
\\
g_\eta(\tau ,\theta ) &=&\frac 1{\cos \theta }\left[ \eta (\tau )- \eta (\tau \sin ^2\theta ) 
\sin\theta \right],  \label{g} 
\end{eqnarray}
for $\tau\in[0,T], \theta\in [0,\frac{\pi}{2}]$. According to \cite{SSC}, the asymptotic analysis of the above integral equation for the unknown function $\eta(\tau)$  enables us to conclude the asymptotic approximation formula for $\eta(\tau)$ as $\tau\to 0$. 
The early exercise behavior of $\varrho(\tau)$ for $\tau\to 0$ can be then deduced from the second order iteration to the system (\ref{eq:SSCintegralequation}) and (\ref{F}) when starting from the initial guess $\eta_0(\tau) = (r-\frac{\sigma^2}{2}) \frac{\tau^{\frac{1}{2}}}{\sigma\sqrt{2}}$ corresponding to the constant early exercise boundary $S_{f0}(t) \equiv E$. One can iteratively compute  $F_{\eta_0}$, $\eta_1$ and $F_{\eta_1}$, $\eta_2$. It turned out from calculation performed in \cite{SSC} that the second consecutive iterate $\eta_2$ is the lowest order (in $\tau$) approximation of $\eta$. Namely, 
\begin{equation}
\eta (\tau )\sim -\sqrt{-\ln \left[ \frac{2 r}{\sigma} \sqrt{2\pi \tau } e^{r \tau
}\right] } \quad \hbox{as}\ \tau\to 0^+.
\label{canad1-eta2}
\end{equation}

Interestingly enough, it has been shown just recently by Chen \emph{et al.} \cite{CCJZ} that the early exercise boundary function $\varrho$ is convex (see also \cite{Ch,CC}). Moreover, the approximation formula (\ref{canad1-eta2}) derived by  Stamicar, \v{S}ev\v{c}ovi\v{c} and Chadam \cite{SSC} provides the right asymptotic behavior for $\tau\to 0^+$. Furthermore, Chen and Chadam \cite{CC} derived sixth-th order expansion of the function 
\begin{equation}
\alpha(\tau) = - \xi - \frac{1}{2\xi} + \frac{1}{8\xi^2} + \frac{17}{24\xi^3}
-\frac{51}{64\xi^4}-\frac{287}{120\xi^5}+\frac{199}{32\xi^6}+O(\xi^{-7}), 
\label{eq:asymptotics}
\end{equation}
for $\xi = \ln {\sqrt{\frac{8\pi r^2 \tau}{\sigma^2}}}\to -\infty$ as $\tau\to 0^+$ where
\begin{equation}
\varrho(\tau) = E e^{ - \sigma \sqrt{2\tau \alpha(\tau)}}.
\label{eq:ro_ssh}
\end{equation}

\subsubsection*{Early exercise boundary asymptotic close to expiry}

Similarly as in the case of the analytic approximation formula by Zhu, we examine the asymptotic behavior of the function $\varrho(\tau)$ for $\tau\to 0$ where $\varrho(\tau)\equiv\varrho^{SSC}(\tau)$ is given by the equation:
\begin{equation}
\varrho (\tau )=Ee^{-(r-\frac{\sigma^2}{2})\tau  + \sigma \sqrt{2 \tau }\tilde\eta (\tau )},\quad
\hbox{where}\ \ \tilde\eta (\tau ) =  -\sqrt{-\ln \left[ \frac{2 r}{\sigma} \sqrt{2\pi \tau } e^{r \tau
}\right] }.
\label{eq:SSC-new}
\end{equation}
Employing expression (\ref{eq:SSC-new}) it is straightforward to verify that 
\[
\lim_{\tau\to 0^+} 
\frac{E-\varrho^{SSC}(\tau)}{\sqrt{\tau}\sqrt{-\ln\tau}} = E \sigma.
\]
Again, as a consequence of the above limit we conclude the following asymptotic approximation of the analytic valuation formula due to Stamicar, \v{S}ev{c}ovi\v{c} and Chadam:
\begin{equation}
\varrho^{SSC}(\tau) \approx E \left(1 - \sigma\sqrt{\tau}\sqrt{-\ln\tau} \right)  \quad \hbox{for} \ 0<\tau\ll 1,
\label{eq:asymptotic-SSC}
\end{equation}
i.e. $\varrho^{SSC}(\tau) = E \left(1 - \sigma\sqrt{\tau}\sqrt{-\ln\tau} \right) + o(\sqrt{\tau}\sqrt{-\ln\tau})$ as $\tau\to0^+$.
Notice that the asymptotic formula (\ref{eq:asymptotic-SSC}) differs from the one obtained from Zhu's formula (\ref{eq:asymptotic-Zhu}) by a logarithmic factor  $\sqrt{-\ln\tau}$.

\subsection{Approximation formulae by Evans, Kuske and Keller}

In \cite{KK} Kuske and Keller proposed another analytic approximation of the early exercise boundary for times close to expiration. Then, in the paper with Evans \cite{EKK}, they improved and extended the formula for the case of dividend-paying asset. 

We begin with the approximation formula by Kuske and Keller \cite{KK}. Their approximation formula for the position of the early exercise boundary close to expiry $t\to T$ reads as follows:
\begin{equation}
\varrho^{KK}(\tau) \approx E\left(
1 -\sigma\sqrt{2\tau}\sqrt{-\ln{ \left[ \frac{2r}{\sigma} \sqrt{\frac{9 \pi \tau}{2} }  \right]   }}
\right),\qquad \hbox{as}\ \ \tau\to0^+.
\label{eq:KK}
\end{equation}
In \cite{EKK} Evans, Kuske and Keller derived an improved asymptotic formula:
\begin{equation}
\varrho^{EKK}(\tau) \approx E\left(
1 -\sigma\sqrt{2\tau}\sqrt{-\ln{\left[    \frac{2r}{\sigma} \sqrt{2\pi\tau}  \right] }}
\right),\qquad \hbox{as}\ \ \tau\to0^+.
\label{eq:EKK}
\end{equation}
Although, asymptotic formulae (\ref{eq:KK}), (\ref{eq:EKK}) by Evans, Kuske and Keller and Stamicar, \v{S}ev\v{c}ovi\v{c} and Chadam (\ref{eq:SSC-new}) differ in higher order terms of $\tau$, it holds
\begin{equation}
\lim_{\tau\to 0^+} 
\frac{E-\varrho^{SSC}(\tau)}{\sqrt{\tau}\sqrt{-\ln\tau}} = \lim_{\tau\to 0^+} 
\frac{E-\varrho^{KK}(\tau)}{\sqrt{\tau}\sqrt{-\ln\tau}} =\lim_{\tau\to 0^+} 
\frac{E-\varrho^{EKK}(\tau)}{\sqrt{\tau}\sqrt{-\ln\tau}} =
E \sigma.
\label{eq:alllim}
\end{equation}
It means that approximation formulae due to Evans, Kuske and Keller \cite{KK,EKK} and Stamicar, \v{S}ev\v{c}ovi\v{c} and Chadam \cite{SSC} have the same asymptotic behavior close to expiry $t\approx T$, i.e. $0<\tau\ll 1$.

\section{Numerical methods for calculation of the early exercise boundary}
\label{sec:numerical}
The  early exercise boundary function $\varrho(\tau)$ for the entire time interval $\tau\in[0,T]$, can be approximated by using numerical methods as well. In this section we present two approaches: 1) a new local iterative algorithm based on the integral equation due to Stamicar, \v{S}ev\v{c}ovi\v{c} and Chadam \cite{SSC}; 2) the well-known PSOR method (c.f. Kwok \cite{Kw}). 

\subsection{A new numerical algorithm based on a solution to the integral equation}
\label{sec:ssch-method}

The aim of this section is to introduce a new  numerical algorithm for computation of the early exercise boundary of the American put option. It is based on a solution to the system of implicit equations (\ref{eq:SSCintegralequation}), (\ref{F}), (\ref{g}) derived by Stamicar \emph{et al.} in \cite{SSC}. The idea of the proposed algorithm is to sequentially compute values of the auxiliary function $\eta=\eta(\tau)$ in nodal points $\tau_i\in [0,T]$. In contrast to global iterative algorithms which iteratively compute the entire solution $\varrho(\tau), \tau\in[0,T],$ (see e.g. \v{S}ev\v{c}ovi\v{c} \cite{Se1}) we only need to find a root of a real valued function at each nodal point $\tau_i$. This is due to the form of functions $F_\eta, g_\eta$ (see (\ref{F}) and (\ref{g})) whose values at $\tau\in (0,T]$ depend only on the value $\eta(\tau)$ and the history path  $\{\eta(\xi), 0\le \xi<\tau\}$. 

Our new algorithm for computation of the approximation of the early exercise boundary $\varrho(\tau)=S_f(T-\tau)$  reads as follows:

\begin{enumerate}

\item{} Construct a division $0=\tau_0 < \tau_1 < ... < \tau_m=T$ of the interval $[0, T]$. 
To this end we can employ either equidistant partition  $\tau_i = (i/m) T$, or we can use $\tau_i = (i/m)^2 T$ in order to adjust the discretization mesh to desired behavior (\ref{eq:asymptotic-SSC}) of $\varrho(\tau)$ close to expiry $\tau\approx 0$. We take $m\gg 1$ sufficiently large such that $\frac{2 r}{\sigma} \sqrt{2\pi \tau_1 } e^{r \tau_1 }<1$.

\item{} Compute the value of $\eta_1\approx \eta(\tau_1)$ from the analytic approximation formula (\ref{canad1-eta2}), i.e. 
\[
\eta_1 = -\sqrt{-\ln \left[ \frac{2 r}{\sigma} \sqrt{2\pi \tau_1 } e^{r \tau_1
}\right] }.
\]

\item{} for $i=2, ..., m$, compute the value $\eta_i\approx \eta(\tau_i)$ as follows:

\begin{itemize}
\item[3-1] Construct the mapping ${\mathcal G}_{\eta_i}(\tau_i, \theta) = \frac 1{\cos \theta }\left[ \eta_i -\tilde\eta (\tau_i \sin ^2\theta ) \sin \theta\right],$
where $\tilde\eta(\tau_i \sin ^2\theta)$ is a linear interpolation function between the points 
$(\tau_j,\eta_j)$ and $(\tau_{j+1},\eta_{j+1})$ if $\tau_j \leq \tau_i \sin^2 \theta < \tau_{j+1}$ for some $1\le j<i$. 
If  $0<\tau_i \sin^2 \theta< \tau_1$ then $\tilde\eta(\tau_i \sin ^2\theta)$ is given by the analytic approximation formula (\ref{canad1-eta2}).

\item[3-2] Construct the mapping ${\mathcal F}_{\eta_i}(\tau_i)$:
\[
{\mathcal F}_{\eta_i}(\tau_i)= 2\int_0^{\pi /2}e^{-r\tau_i \cos ^2\theta 
- {\mathcal G}^2_{\eta_i}(\tau_i ,\theta)}
\left( 
\frac{\sigma\sqrt{\tau_i }}{\sqrt{2}}\sin \theta +{\mathcal G}_{\eta_i}(\tau_i ,\theta )\tan \theta 
\right)
\,d\theta .
\]
As for the numerical quadrature of the above integral we can employ the composed Newton-Cotes method of the fourth order with, at least, 1000 subintervals. 

\item[3-3] Find the root $\eta_i$ of the equation:
\[
\eta_i =-\sqrt{-\ln \left[ \frac{r\sqrt{2 \pi \tau_i}}{\sigma} e^{ r\tau_i }\left(
1-\frac{{\mathcal F}_{\eta_i}(\tau_i )}{\sqrt{\pi }}\right) \right] }.
\]
The above equation can be solved using either bisection method, Newton's method or any other numerical iterative method for finding roots of real valued functions. In order to speed-up convergence we can use already constructed value $\eta_{i-1}$ as a starting point for iterations at the time level $\tau_i$.

\end{itemize}

\item{} Go to step 3 and repeat the calculation of $\eta_i$ for the next value of $i$ until $i\le m$.

\item{}
From discrete values $\eta_i, i=1,2, ..., m,$ we compute the approximation $\varrho_i$ of the early exercise boundary position $\varrho(\tau_i)$ as follows: 
\[
\varrho_i = E e^{-(r-\frac{\sigma^2}{2})\tau_i  + \sigma \sqrt{2 \tau_i }\eta_i},
\]
We set $\varrho_0= E$. 
The entire profile $\varrho(\tau)=S_f(T-\tau), \tau\in[0,T]$, is then computed as a linear interpolation function between discrete values $\{ (\tau_i,\varrho_i), i=0, ..., m\}$.

\end{enumerate}

\subsection{Approximate solution using the PSOR method}
\label{sec:psor}

In this section, we present a brief overview how the early exercise boundary can be found using a finite difference numerical approximation method applied to the variational inequality (\ref{var-amer-put}). The method consits in computation the option price $V(S,t)$ using the so-called projected successive over relaxation (PSOR) method introduced by Ockendon and Elliot in \cite{EO}.
Having computed a solution $V(S,t)$ to the variational inequality (\ref{var-amer-put}) we can calculate the early exercise boundary position. Indeed, given a time $t$, the critical stock price $S_f(t)$ is equal to the maximal stock price $S=S_f(t)$ for which the option price  is equal to the payoff, i.e. 
\[
S_f(t) = \max \{ S>0 \ | \ V(S,t) =  (E-S)^+ \}.
\]
Following Kwok \cite{Kw}, the idea of the PSOR method is to transform (\ref{var-amer-put}) by introducing new variables $x=\ln(S/E), \quad \tau=T-t, \quad u(x,\tau) = e^{\alpha x + \beta \tau } V(E e^x, T-\tau)$
where constants $\alpha,\beta$ are defined by $\alpha = \frac{r}{\sigma^2} -\frac{1}{2}, \beta = \frac{r}{2}+ \frac{\sigma^2}{8} + \frac{r^2}{2\sigma^2}$. 
We denote $u_i^j \approx u(i h, j k)$ the finite difference approximation of a solution to the transformed variational inequality for $i=-n, ..., -1,0,1, .., n$, $j=1, ..., m$. The spatial and time discretization 
steps $h, k > 0$ are chosen such that $h=L/n$, $k=T/m$, respectively. Here $T$ represents expiration time and $L$ is a sufficiently large bound for the interval $x\in (-L,L)$. For practical purposes we can take $L\approx 1$ .
In each time step $j=1,2, ..., m$, a linear complementarity problem for the finite difference approximation vector $u^j\in \R^{2n+1}$ is solved by using the iterative successive over relaxation (SOR) method where iterates are projected to the transformed pay-off diagram. This is done by taking the maximum of the transformed pay-off and computed iteration of a solution obtained by the SOR successive iteration. For details we refer the reader to \cite[pp. 212--224]{Kw}.

\section{Numerical comparison of the early exercise boundary approximations}
\label{sec:comparison}

\subsection{Comparison of approximations close to expiry}

\begin{figure}
\begin{center}
\subfigure[$T = 5\times10^{-5}$  (1 min)]{
  \includegraphics[height=4.5cm]{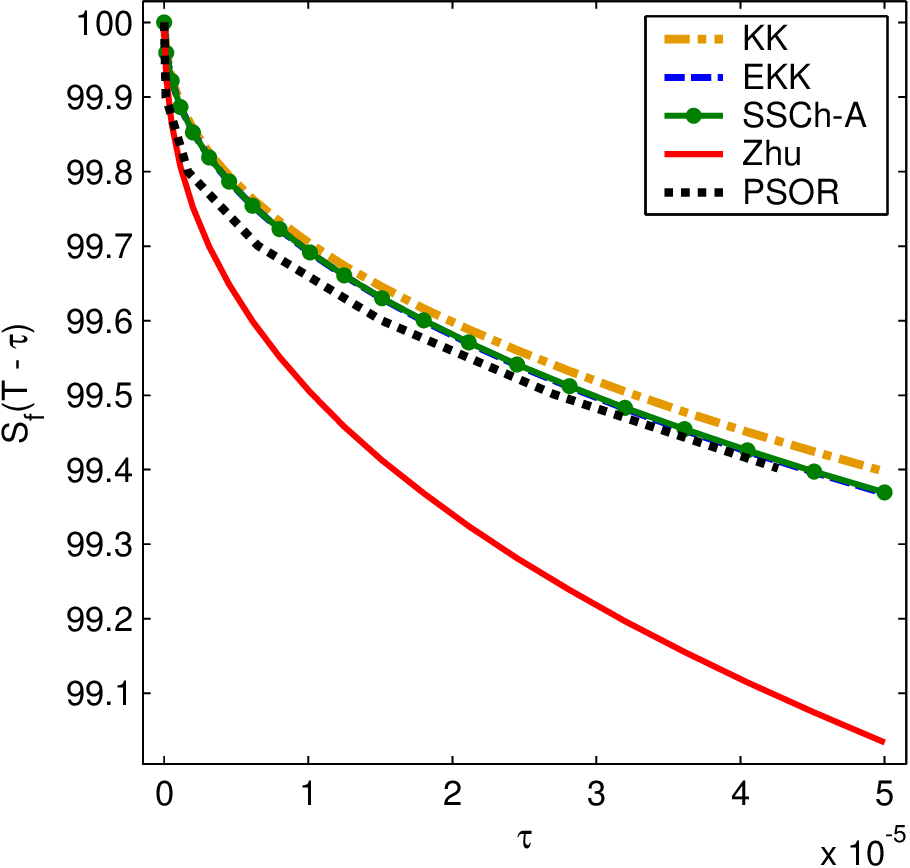}
}
\subfigure[$T = 4\times 10^{-3}$ (1 day)]{
  \includegraphics[height=4.5cm]{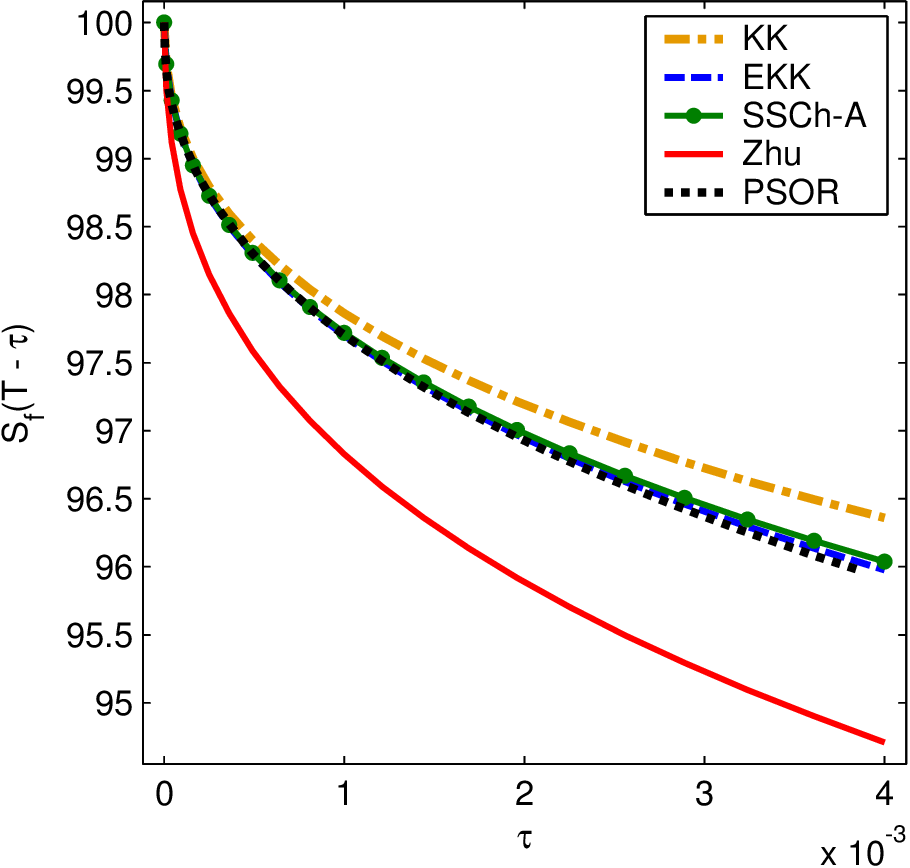}
}
\subfigure[$T = 0.08 $  (1 month)]{
  \includegraphics[height=4.5cm]{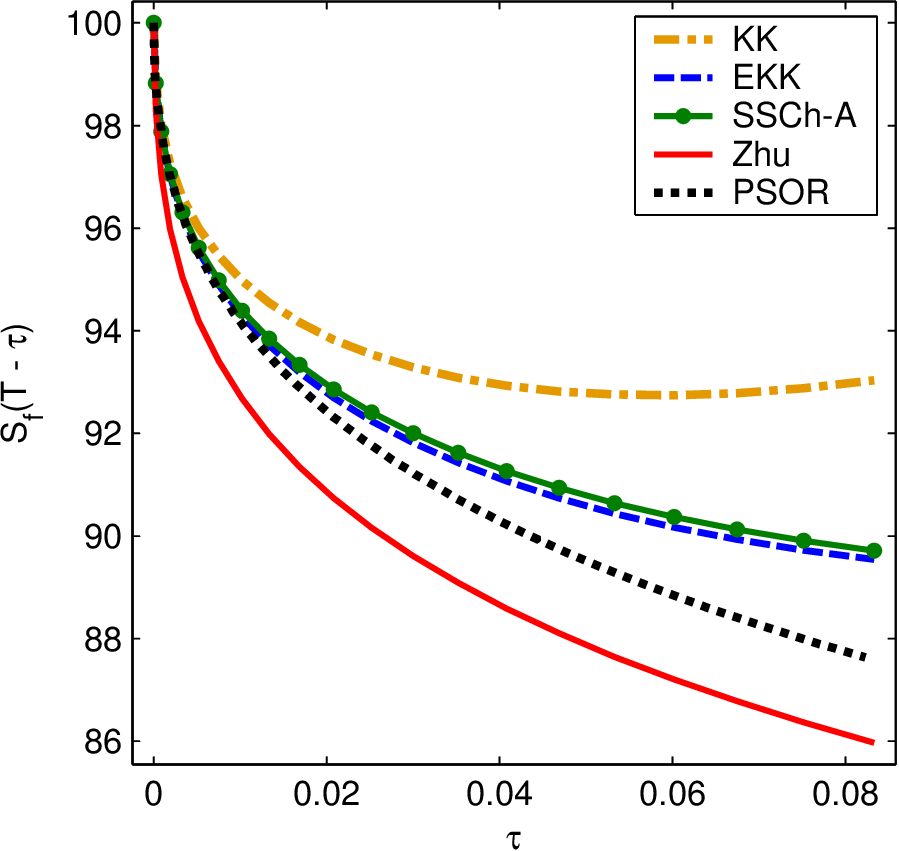}
}
\subfigure[$T = 0.25 $ (3 months)]{
  \includegraphics[height=4.5cm]{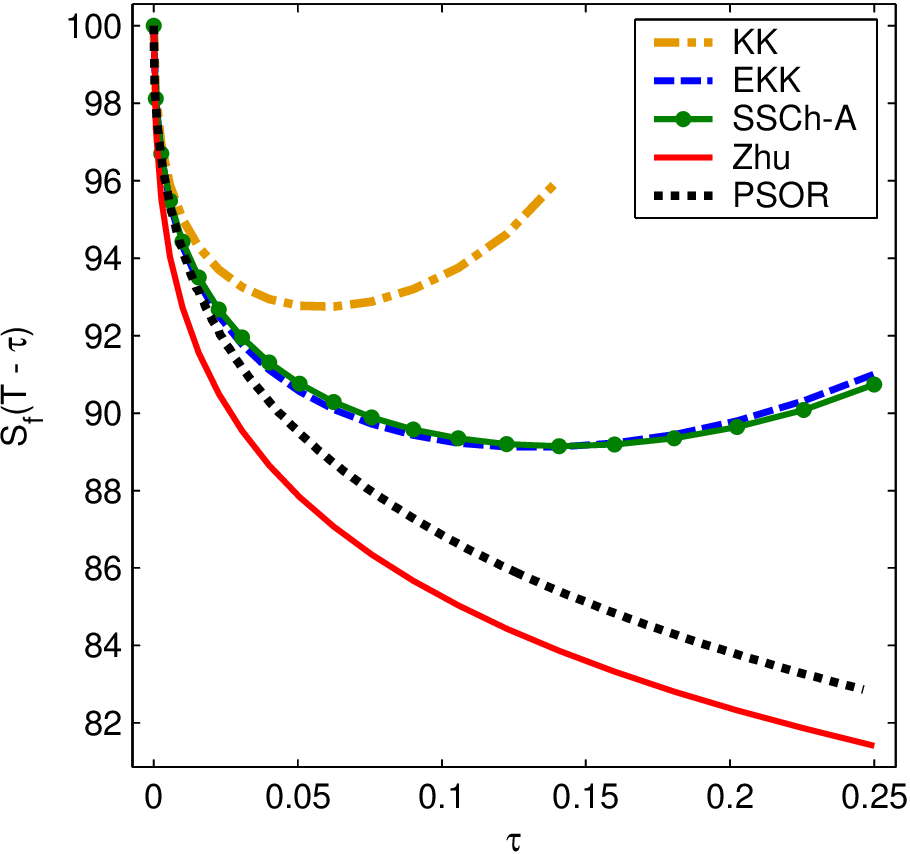}
}
\end{center}

\caption{Comparison of analytic approximation formulae for various maturities $T$ on a yearly basis.}

\label{fig:analyticke}
\end{figure}

This section focuses on numerical comparison of analytic approximations due to Stamicar, \v{S}ev\v{c}ovi\v{c} and Chadam (\ref{eq:SSC}), Kuske and Keller (\ref{eq:KK}), Evans, Kuske and Keller (\ref{eq:EKK}), Zhu (\ref{eq:zhu}) 
and our new local iterative algorithm from section 3.1 for the early exercise boundary for times $0<\tau=T-t\ll 1$ close to expiry. 
For computational purposes we chose the volatility $\sigma=30\%$, risk-free interest rate $r=10\%$ p.a., and the strike price $E=100\$$. 

\begin{table}
\caption{\small
Comparison of the early exercise boundary obtained by analytic approximation formulae and the iterative algorithm to the benchmark PSOR method.}
\begin{center}
\scriptsize
\begin{tabular}{l|ccccc|cccc} \hline 
\multirow{2}{*}&  \multicolumn{5}{|c|}{\bf Early exercise boundary}   &   \multicolumn{4}{|c}{\bf Relative error}  \\ 
\multirow{2}{*} {$\tau$}&  \multicolumn{5}{|c|}{$\varrho(\tau)=S_f(T-\tau)$ }   &   \multicolumn{4}{|c}{\bf w.r. to the PSOR method}  \\ 
& {\bf  EKK }  &  {\bf Zhu } &  {\bf SSCh-A } & {\bf  SSCh  }  &  {\bf PSOR   }  & {\bf  EKK  } &   {\bf Zhu } & {\bf SSCh-A } &  {\bf SSCh  }  \\ \hline\hline
0.000 01 & 99.69 & 99.51 & 99.69 & 99.690 & 99.7 & 0.01\% & 0.19\% & 0.01\% & 0.01\% \\ 
0.000 05 & 99.37 & 99.03 & 99.37 & 99.358 & 99.4 & 0.03\% & 0.37\% & 0.03\% & 0.04\% \\ 
0.000 1 & 99.14 & 98.72 & 99.15 & 99.111 & 99.2 & 0.06\% & 0.49\% & 0.06\% & 0.09\% \\ 
0.000 5 & 98.28 & 97.57 & 98.29 & 98.270 & 98.31 & 0.03\% & 0.76\% & 0.02\% & 0.04\% \\ 
0.001 & 97.7 & 96.83 & 97.72 & 97.660 & 97.73 & 0.03\% & 0.92\% & 0.01\% & 0.07\% \\ 
0.01 & 95.62 & 94.27 & 95.69 & 95.502 & 95.6 & 0.02\% & 1.39\% & 0.09\% & 0.10\% \\ 
0.01 & 94.33 & 92.73 & 94.43 & 94.070 & 94.18 & 0.16\% & 1.54\% & 0.27\% & 0.11\% \\ 
0.04 & 91.12 & 88.66 & 91.31 & 90.205 & 90.3 & 0.90\% & 1.82\% & 1.12\% & 0.11\% \\ 
0.1 & 89.29 & 85.25 & 89.42 & 86.762 & 86.94 & 2.70\% & 1.93\% & 2.86\% & 0.20\% \\
\hline
\end{tabular}
\end{center}
{\scriptsize 
Legend: EKK - Evans, Kuske and Keller \cite{EKK}, 
SSCh-A - Stamicar, \v{S}ev\v{c}ovi\v{c} and Chadam \cite{SSC}, 
SSCh - our new local iterative algorithm from section 3.1,
ZHU - Zhu \cite{Zhu2006}, PSOR - Projected SOR method \cite{Kw}. 
}
\label{tab:analyticke}
\end{table}

In Fig.~\ref{fig:analyticke} we present quantitative comparison of analytic approximation formulae by Kuske and Keller (KK),  Evans, Kuske and Keller (EKK), Stamicar, \v{S}ev\v{c}ovi\v{c} and Chadam (SSCh-A), Zhu's formula (ZHU). As a numerical benchmark solution we chose the PSOR method  with $n=1000$ spatial grid points and $m=1000$ time steps (see section \ref{sec:psor}). It should be obvious that the approximation formulae KK, EKK and SSCh-A exhibit similar behavior w.r. to PSOR for the time close to expiry (see Fig.~\ref{fig:analyticke} a), b)). On the other hand, on a larger time horizon KK, EKK as well as SSCh-A become nondecreasing and Zhu's formula (ZHU) better approximates the PSOR solution (see Fig.~\ref{fig:analyticke} c), d)). It is also worthwile to note that Zhu's formula undershoots the early exercise boundary for small values of $\tau$ when compared to KK, EKK, SSCh-A and PSOR. This phenomenon can be easily justified by calculating the limit
\begin{equation}
\lim_{\tau\to 0^+} 
\frac{E-\varrho^{SSC}(\tau)}{E-\varrho^{Zhu}(\tau)} \sqrt{-\ln\tau}
= 1.
\label{eq:undershoot}
\end{equation}

In Table~\ref{tab:analyticke} we calculated the early exercise boundary position for EKK, ZHU, SSCh-A, PSOR methods and our new local iterative algorithm described in section 3.1 which is labeled as SSCh. 
We also calculated the relative error $\Delta^{method}(\tau)$ defined as 
\[
\Delta^{method}(\tau) = \frac{\left|S_f^{method}(T-\tau)-S_f^{PSOR}(T-\tau)\right|}{S_f^{PSOR}(T-\tau)}, \quad \hbox{for} \ \tau\in [0,T],
\]
where $S_f^{PSOR}$ is the early exercise boundary computed by the PSOR method. For $\tau \approx 1$ minute  EKK, SSCh-A and PSOR methods have almost identical values (close to $99.4\$$) whereas Zhu's boundary position has been  calculated as $99.03\$$. On the other hand, other approximations (EKK and SSCh-A) differs significantly from early exercise boundary obtained by the PSOR method as we enlarge time to expiration $\tau>0.02$. The relative error in the early exercise boundary position calculated by Zhu's formula w.r. the PSOR method is less than $2\%$. The best approximation of the early exercise boundary has been achieved by our local iterative algorithm SSCh.

In summary, SSCh-A, KK and EKK analytic approximation formulae are suitable for  approximation of the early exercise boundary close to expiration whereas, for a longer time horizon, it is recommended to use the analytical approximation formula by Zhu. The new local iterative approximation derived in section 3.1 can be used for both small as well as large time horizon.

\begin{table}

\caption{\small Comparison of the early exercise boundary on a long time horizon.}

\begin{center}

\scriptsize
\begin{tabular}{l | ccc | cc} \hline

\multirow{2}{*}{ $\tau$ } 
&  
\multicolumn{3}{|c|}{ \bf Early exercise boundary } & \multicolumn{2}{|c}{ \bf Rel. error  w.r. to}  \\ 
&  
\multicolumn{3}{|c|}{  $S_f(T-\tau)$} & \multicolumn{2}{|c}{ \bf PSOR method}  \\ 
&   {\bf PSOR}     &  {\bf Zhu }&  {\bf SSCh}  & { \bf Zhu}  &  {\bf SSCh}   \\ 
\hline \hline
0 & 100. & 100. & 100. & 0\% & 0\% \\
0.02 & 92.8672 & 90.8575 & 92.3461 & 2.16\% & 0.56\% \\
0.04 & 90.7707 & 88.6563 & 90.2088 & 2.33\% & 0.62\% \\
0.06 & 89.3300 &   87.2160 &  88.7771 &  2.37\% &   0.62\% \\
0.08 & 88.2350 & 86.1300 & 87.6695 & 2.39\% & 0.64\% \\
0.1 & 87.3279 & 85.2538 & 86.7636 & 2.38\% & 0.65\% \\
0.2 & 84.2962 & 82.3766 & 83.7476 & 2.28\% & 0.65\% \\
0.4 & 81.0179 & 79.3593 & 80.4793 & 2.05\% & 0.66\% \\
0.6 & 79.0571 & 77.5961 & 78.5391 & 1.85\% & 0.66\% \\
0.8 & 77.6986 & 76.3752 & 77.1895 & 1.7\% & 0.66\% \\
1 & 76.6695 & 75.4580 & 76.1632 & 1.58\% & 0.66\% \\
1.5 & 74.9137 & 73.8879 & 74.4094 & 1.37\% & 0.67\% \\
2 & 73.8107 & 72.8731 & 73.2722 & 1.27\% & 0.73\% \\
3 & 72.5786 & 71.6205 & 71.8735 & 1.32\% & 0.97\% \\
4 & 72.0121 & 70.8778 & 71.0464 & 1.58\% & 1.34\% \\
5 & 71.7966 & 70.3925 & 70.5100 & 1.96\% & 1.79\% \\
 \hline
\end{tabular}
\end{center}

{\scriptsize 
Legend: 
SSCh - our new local iterative algorithm from section 3.1, ZHU - Zhu \cite{Zhu2006}, PSOR - \cite{Kw}.
}
\label{tab:comparison}
\end{table}

\subsection{The long term horizon}

In the long term horizon, i.e. $\tau=T-t \approx 1 $ year or even more, we can no longer use the analytical  approximations SSCh-A, KK, EKK designed for $0<\tau\ll 1$ any more. These solutions loose monotonicity for $\tau \approx 0.1$ and they become even undefined for large values of $\tau$ because of the sign change in the logarithm. 
This is why only Zhu's analytical approximation formula for the early exercise boundary (\ref{eq:zhu}) can be used in the long term horizon. We compared Zhu's approximation with two numerical methods described in section \ref{sec:numerical}. The first method is our new numerical method (labeled by SSCh) based on the integral equation (\ref{eq:SSCintegralequation}) which was described in section~\ref{sec:ssch-method}. The second method is the classical PSOR method described in section~\ref{sec:psor}. 

\begin{figure}
\begin{center}
  \includegraphics[height=4.5cm]{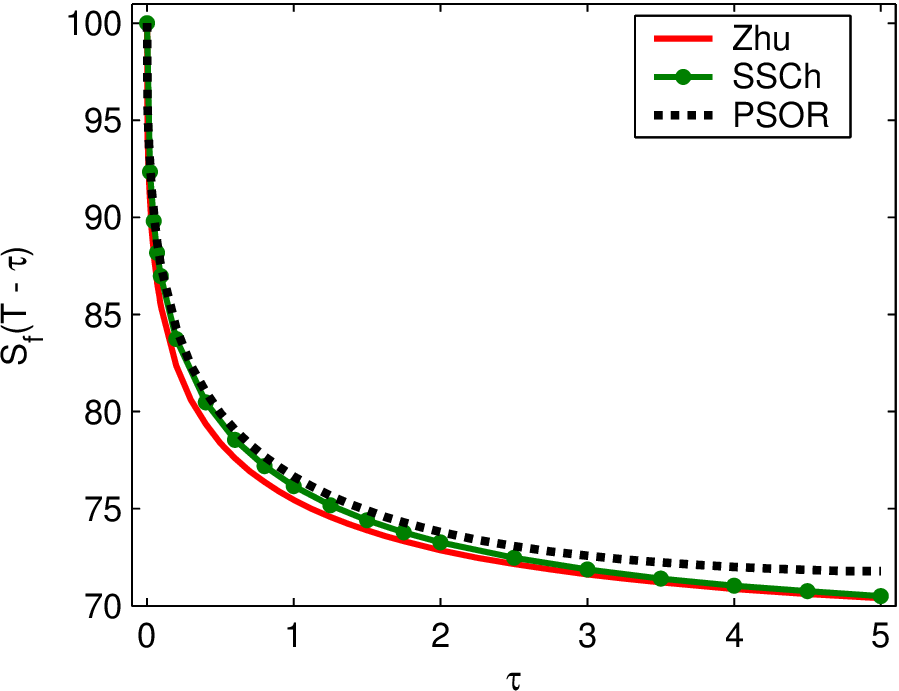}
\hskip 5truemm
  \includegraphics[height=4.5cm]{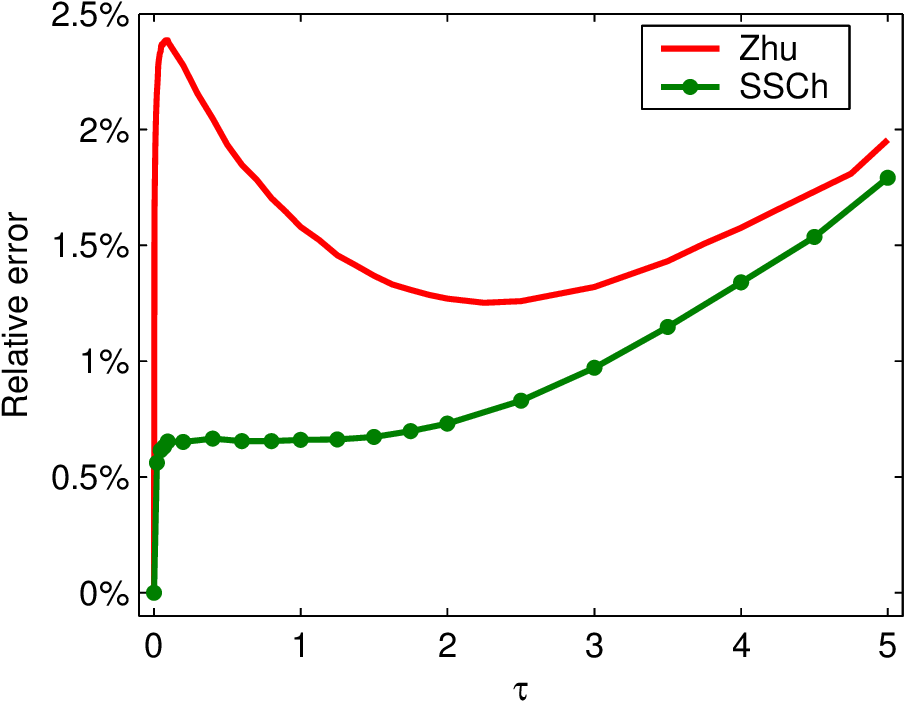}
\end{center}

\caption{\small Comparison of the early exercise boundary position in the long time horizon.
The early exercise boundary position (left). The relative error with respect to the PSOR method (right).
}
\label{fig:years5}
\end{figure}

As a time horizon, we chose the large expiration time $T=5$ years. Other model parameters are the same as in the previous section, i.e. $E=100\$ $, $\sigma=30\%$,  $r=10\%$ p.a. 
The computational results are shown in Fig.~\ref{fig:years5} (left) and Table~\ref{tab:comparison}. We can observe that the shape of all solutions is very similar. 
In Fig.~\ref{fig:years5} (right) we plotted the relative error with respect to the PSOR method, which was used as a benchmark. Zhu's analytic approximation formula has the relative error between $1$ and $2.5\%$ and it attained local minimum around $\tau \approx 2$ years. This is due to the fact that Zhu's method is slightly undershooting the solution close to expiry, i.e. for $\tau\approx 0$. 
The solution computed by our new SSCh scheme shows nearly constant error term until $\tau \approx 2.5$ years, then the relative error starts to grow up. For $\tau\approx 5$ years, the numerical solution SSCh is approaching Zhu's approximation. This is due to loose of precision of the PSOR method itself when the exact early exercise boundary could be closer to SSCh and Zhu's approximation  than to the PSOR solution.

\setcounter{subfigure}{0}

\section{Comparison of options prices}

In this section we address the question concerning the difference between the American put option price and the approximative option price computed with an approximation of the early exercise boundary. More precisely, let $V^{am}(S,t)$ be the solution to the free boundary problem (\ref{amer-put}) with the early exercise boundary profile $S_f$. Let us consider a given function $S^{app}_f$ representing an approximation of the early exercise boundary profile $S_f$. We denote by $V^{app}$ the unique solution to the parabolic equation:

\begin{figure}
\begin{center}
\includegraphics[width=7.5cm]{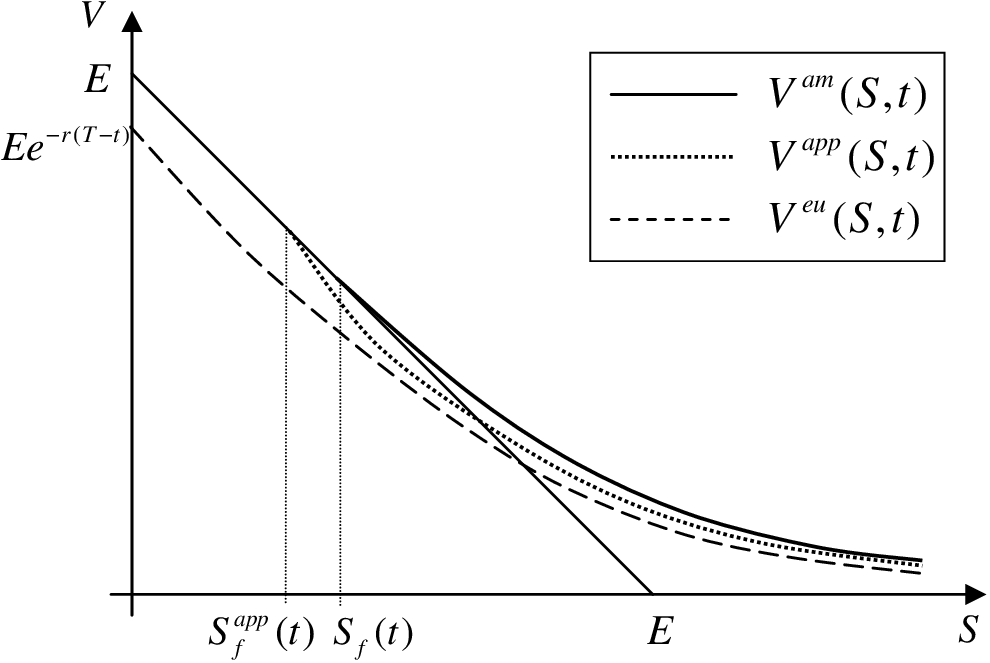}
\end{center}
\caption{A profile $S\mapsto V^{am}(S,t)$ of the American option price and its comparison to the option price $V^{app}$ computed with respect to the approximative early exercise boundary $S^{app}_f(t)=\varrho^{app}(T-t)$. The corresponding European style of an option is labeled by $V^{eu}$. }
\label{fig:comparison}
\end{figure}

\begin{eqnarray}
&&\frac{\partial V^{app}}{\partial t} +  r S\frac{\partial V^{app}}{\partial S} + {\sigma^2\over 2} S^2 \frac{\partial^2 V^{app}}{\partial S^2} - r V^{app} =0\,,
\quad t\in(0,T),\ S^{app}_f(t) < S \,,
\nonumber
\\
&&V^{app}(+\infty ,t)=0,\ V^{app}(S^{app}_f(t), t)= E - S^{app}_f(t)\,,
\\
&&V^{app}(S,T)=(E-S)^+\,.
\nonumber
\label{amer-put-appr}
\end{eqnarray}
Notice that we do not require the solution $V^{app}$ to satisfy the $C^1$ smooth pasting contact condition $ \frac{\partial V^{app}}{\partial S}(S^{app}_f(t),t)=-1$. In fact,  $V^{app}$ is a solution to the put barrier option (cf. Kwok \cite{Kw}) with a given down-and-out barrier $t\mapsto S^{app}(t)$. For asset prices $0<S<S^{app}_f(t)$ we set
\[
V^{app}(S,t) = E- S. 
\]
A comparison of the profile $S\mapsto V^{am}(S,t)$ of the American option price and the approximative option price $V^{app}$ is shown in Fig.~\ref{fig:comparison}. We also plot the common lower bound for both put option prices represented by the plain vanilla European put option labeled by $V^{eu}$.

Knowing the functions $t\mapsto S_f(t)$ and  $t\mapsto S^{app}_f(t)$ it is not difficult to calculate the difference  $V^{am}(S,t) - V^{app}(S,t)$ between option prices. Indeed, using the  standard transformation (see e.g. Kwok \cite{Kw})
\[
V^{am}(S,t) = E e^{\alpha x+\beta \tau} u^{am}(x,\tau), \quad 
V^{app}(S,t) = E e^{\alpha x+\beta \tau} u^{app}(x,\tau), 
\]
where
\[
x=\ln(S/E), \ \tau=T-t, \qquad \alpha=\frac{1}{2} -\frac{r}{\sigma^2}, \quad  \beta = -\frac{r}{2} - \frac{r^2}{2\sigma^2} - \frac{\sigma^2}{8},
\]
taking into account the fact that $V^{am}(S,t) = E-S$ for $0<S<S_f(t)$ and $V^{app}(S,t) = E-S$ for $0<S<S^{app}_f(t)$ we can conclude that $u^{am}, u^{app}$ are solution to the following Cauchy problems:
\[
\frac{\partial u^{am}}{\partial \tau} - \frac{\sigma^2}{2} \frac{\partial^2 u^{am}}{\partial x^2} = 
 \left\{
\begin{array}{cc}
0& \hbox{for}\ x> \ln(\varrho(\tau)/E),  \\ 
r e^{-\alpha x-\beta \tau} & \hbox{for}\ x\le \ln(\varrho(\tau)/E),
\end{array} 
 \right. 
\]
\[
\frac{\partial u^{app}}{\partial \tau} - \frac{\sigma^2}{2} \frac{\partial^2 u^{app}}{\partial x^2} = 
\left\{
\begin{array}{cc}
0& \hbox{for}\ x> \ln(\varrho^{app}(\tau)/E),  \\ 
r e^{-\alpha x-\beta \tau} & \hbox{for}\ x\le \ln(\varrho^{app}(\tau)/E),
\end{array} 
 \right. 
\]
defined for $-\infty<x<\infty, 0<\tau<T,$ where $\varrho(\tau)=S_f(T-\tau), \varrho^{app}(\tau)=S^{app}_f(T-\tau)$. Notice that the difference $v(x,\tau)=u^{am}(x,\tau)-u^{app}(x,\tau)$ satisfies $v(x,0)=0$ for each $x\in\R$. Using Green's representation formula for a solution to a linear parabolic equation we obtain, after some calculations, the explicit expression for the difference of option prices
\begin{eqnarray}
&&V^{am}(S,t) - V^{app}(S,t) 
\\
&& 
\quad 
= 
r E \int_0^{\tau} \left| \int_{\ln(\varrho^{app}(\xi)/E)}^{\ln(\varrho(\xi)/E)}
G(x - s, \tau -\xi )e^{\alpha (x-s)+\beta (\tau-\xi)} ds \right| d\xi,
\nonumber 
\label{difference-prices}
\end{eqnarray}
where $G(x,\tau) =  e^{-x^2/(2\sigma^2\tau)}/\sqrt{2\pi\sigma^2\tau}$ is the Green function. The above difference in option prices is always nonnegative because the American option price is greater or equal to the price of a down-and-out barrier option with the prescribed barrier $S^{app}_f(t)=\varrho^{app}(T-t)$ (see Kwok \cite{Kw}). 

If we evaluate this difference at the American option early exercise boundary position $S_f(t)$ then we obtain a slightly simplified expression:
\begin{eqnarray}
&&V^{am}(S_f(t),t) - V^{app}(S_f(t),t) 
\\
&& \quad = 
r E \int_0^{\tau} e^{-r (\tau-\xi)}
\left|
N(\tilde\gamma(\tau,\xi)) - N(\gamma(\tau,\xi))
\right|\, d\xi \nonumber
\label{difference}
\end{eqnarray}
where $\tau=T-t$ and 
\[
\tilde \gamma(\tau,\xi) = \frac{\ln\frac{\varrho(\tau)}{\varrho^{app}(\xi)} + (r-\sigma^2/2) (\tau-\xi)}{\sigma\sqrt{\tau-\xi}},
\quad
\gamma(\tau,\xi) = \frac{\ln\frac{\varrho(\tau)}{\varrho(\xi)} + (r-\sigma^2/2) (\tau-\xi) }{\sigma\sqrt{\tau-\xi}}.
\]
Notice that the difference $V^{am}(S,t) - V^{eu}(S,t)$ of the American and European style of put options is rather small. Similarly, the difference  $V^{am}(S,t) - V^{app}(S,t)$ is small. Therefore it is reasonable to calculate the mispricing error $V^{am}(S,t) - V^{app}(S,t)$ with respect to the benchmark mispricing difference $V^{am}(S,t) - V^{eu}(S,t)$ evaluated at $S=S_f(t)$. To this end, let us introduce the following relative mispricing error function:
\begin{equation}
err(T-t) = \frac{V^{am}(S_f(t),t) - V^{app}(S_f(t),t)}{V^{am}(S_f(t),t) - V^{eu}(S_f(t),t)}.
\label{error}
\end{equation}
The denominator of (\ref{error}) can be easily calculated by recalling that 
\[
V^{am}(S_f(t),t)= E  - S_f(t)
\ \ \hbox{and}\ \ 
V^{eu}(S,t)= E e^{-r(T-t)} N(-d_2) - S N(-d_1),
\]
where 
\[
d_1 = \frac{\ln\frac{S}{E} + (r+\sigma^2/2)(T-t) }{\sigma\sqrt{T-t}}, \quad
d_2 = \frac{\ln\frac{S}{E} + (r-\sigma^2/2)(T-t) }{\sigma\sqrt{T-t}}
\]
(see Kwok \cite{Kw}). 

\begin{figure}
\begin{center}
\includegraphics[width=6.5cm]{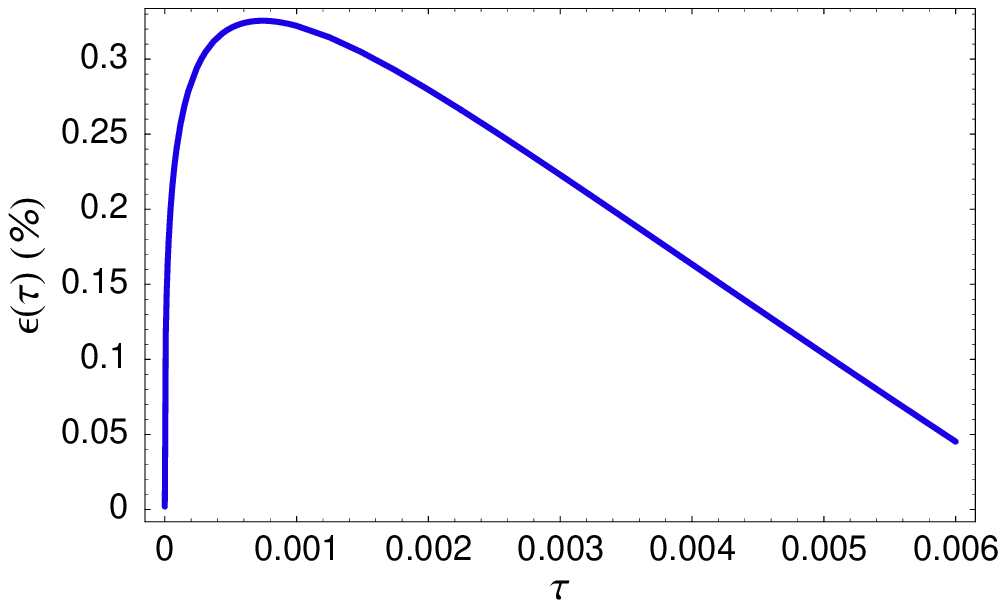}
\includegraphics[width=6.5cm]{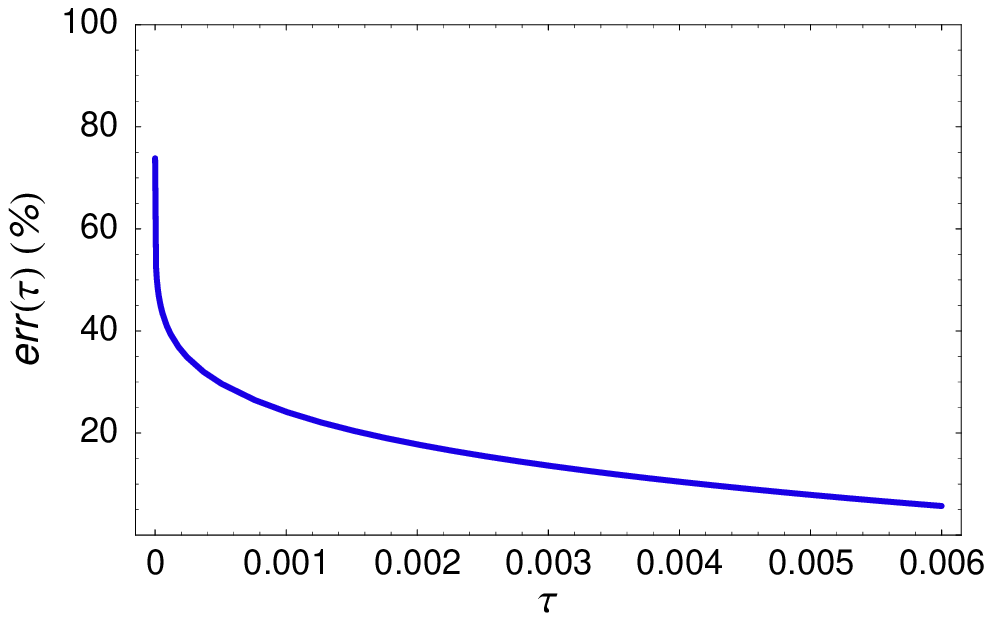}
\end{center}
\caption{Comparison of the early exercise boundary $\varrho(\tau)$ and the approximative early exercise boundary $\varrho^{app}\equiv\varrho^{Zhu}$ obtained from Zhu's formula. The model parameters were chosen as: $E = 1, r = 0.1, \sigma = 0.3$ for the time $\tau=T-t\in(0, 0.006)$ close to expiration.}
\label{fig:comparison2}
\end{figure}

In our practical experiment, we evaluated the relative misspricing error function $err(\tau)$ for the approximation of the early exercise boundary obtained by Zhu, i.e. we set $\varrho^{app}\equiv\varrho^{Zhu}$. In Fig.~\ref{fig:comparison2} (left) we plotted the relative error $\epsilon(\tau)$ in the early exercise boundary position 
\[
\epsilon(T-t) = \frac{S_f(t)-S^{Zhu}_f(t)}{S_f(t)}
\]
between the true early exercise position $S_f(t)=\varrho(T-t)$ and Zhu's approximation $S^{Zhu}_f(t) =  \varrho^{Zhu}(T-t)$. We can see that the maximal relative error  in the early exercise boundary position is only $0.32\%$ and it is  attained six hours prior expiration. 

The relative  error function $err(\tau)$ for times $\tau=T-t$ close to expiry (less than two days) is depicted in Fig.~\ref{fig:comparison2} (right). We can see that the error rapidly increases when the time $t$ approaches expiration $T$. For one day to expiration ($\tau=4\times 10^{-3}$) the error is $15\%$. It increases beyond $70\%$ as $t\to T$. This is due to the fact that Zhu's approximation underestimates the free boundary potion as $\tau=T-t\to 0^+$ (see (\ref{eq:undershoot}) ).

\section{Conclusions}
We presented qualitative and quantitative comparison of analytical approximations and numerical methods for computation the early exercise boundary position of the American put option paying zero dividends. We also proposed a new local iterative numerical scheme for construction of the entire early exercise boundary which is based on a solution to a nonlinear integral equation. We derived asymptotic behavior of approximation formulae for the time close to expiry. We proved that the asymptotic formulae by Evans, Kuske and Keller \cite{KK,EKK},  Stamicar, \v{S}ev\v{c}ovi\v{c} and Chadam \cite{SSC} have the same asymptotic behavior close to expiry. We also showed that the analytic approximation formula by Zhu  \cite{Zhu2006} has a different asymptotic behavior. On the other hand, for a long time horizon, Zhu's formula yields quantitatively the same results as those of our  new local iterative numerical scheme and the numerical benchmark  PSOR method.

\paragraph{Acknowledgments:}
We thank the anonymous referees for their valuable comments and suggestions. This research was supported by the bilateral Slovak--Bulgarian  project APVV SK-BG-0034-08.



\begin{thebibliography}{99}

\small

\bibitem{A}
G.~Alobaidi, R.~Mallier and S.~Deakin, 
\emph{Laplace transforms and installment options},
Math. Models and Methods in Appl. Science
\textbf{18}(8)
(2004),
1167--1189.


\bibitem{AE1}
J.~Ankudinova and M.~Ehrhardt,
\emph{On the numerical solution of nonlinear Black-Scholes equations},
Computers and Mathematics with Applications
\textbf{56}(3)
(2008),
799--812.

\bibitem{BW}
B.~Barone-Adesi and  R.~E.~Whaley, 
\emph{Efficient analytic approximations of American option values},
J. Finance
\textbf{42} 
(1987),
301--320.

\bibitem {BS}
F.~Black and M.~Scholes,
\emph{The pricing of options and corporate liabilities},
J. Political Economy
\textbf{81}
(1973),
637--654.

\bibitem{BJ} 
D.~S.~Bunch and H.~Johnson,
\emph{The American Put Option and Its Critical Stock Price},
The Journal of Finance
\textbf{55}(5)
(2000),
2333--2356.



\bibitem{CJM}
P.~Carr, R.~Jarrow and R.~Myneni,
\emph{Alternative characterizations of American put options},
Mathematical Finance
\textbf{2}
(1992),
87--105.


\bibitem {Ch}
J.~Chadam,
\emph{Free Boundary Problems in Mathematical Finance},
Progress in Industrial Mathematics at ECMI 2006,
Springer Berlin Heidelberg 
\textbf{12} (2008), 655--665.

\bibitem {CCJZ}
X.~Chen, J.~Chadam, L.~Jiang and W.~Zheng,
\emph{Convexity of the Exercise Boundary of the American Put Option on a Zero Dividend Asset},
Mathematical Finance
\textbf{18}(1)
(2008),
185--197.


\bibitem{CC}
X.~Chen and J.~Chadam,
\emph{A mathematical analysis of the optimal exercise boundary for 
American put options},
SIAM J. Math. Anal.
\textbf{38}(5) 
(2007),
1613--1641.



\bibitem{DH}
J.N.~Dewynne, S.D.~Howison, J.~Rupf and P.~Wilmott, 
\emph{Some mathematical results in the pricing of American options}, 
Euro. J. Appl. Math.
\textbf{4} 
(1993),
381--398.



\bibitem{EM}
M.~Ehrhardt and P.~Mickens,
\emph{A fast, stable and accurate numerical method for the Black-Scholes equation of American options}, 
International Journal of Theoretical and Applied Finance,
\textbf{11}(5)
(2008),
471--501.

\bibitem{ET}
E.~Ekstr\"om and  J.~Tysk,
\emph{The American put is log-concave in the log-price},
Journal of Mathematical Analysis and Appl.
\textbf{314}(2) 
(2006),
710--723.

\bibitem {E}
E.~Ekstr\"om, 
\emph{Convexity of the optimal stopping boundary for the American put option},
Journal of Mathematical Analysis and Appl.
\textbf{299}(1)
(2004),
147--156.



\bibitem{EO}
C.~M.~Elliott and J.~R.~Ockendom,
\emph{Weak and Variational Methods for Free and Moving Boundary Problems},
Pitman (1982).


\bibitem{EKK}
J.~D.~Evans, R.~Kuske and J.~B.~Keller,
\emph{American options on assets with dividends near expiry}, 
Mathematical Finance
\textbf{12}
(2002),
219--237.


\bibitem{GJ}
R.~Geske and H.E.~Johnson,  
\emph{The American put option valued analytically},
J. Finance
\textbf{39} 
(1984),
1511--1524.

\bibitem{GR}
R.~Geske and R.~Roll, 
\emph{On valuing American call options with the Black--Scholes European formula},
J. Finance
\textbf{89}
(1984),
443--455.



\bibitem{H}
J.~Hull,
\emph{Options, Futures and Other Derivative Securities},
third edition, 
Prentice-Hall (1997).



\bibitem{J}
H.~Johnson,
\emph{An analytic approximation of the American put price},
J. Finan. Quant. Anal.
\textbf{18}
(1983), 
141--148.


\bibitem{K}
C.~Knessl, 
\emph{A note on a moving boundary problem arising in the American put option}, 
Studies in Applied Mathematics
\textbf{107}
(2001), 
157--183.


\bibitem{K1}
I.~Karatzas,
\emph{On the pricing American options},
Appl. Math. Optim.
\textbf{17}
(1988), 
37--60.


\bibitem{KK}
R.~A.~Kuske and J.~B.~Keller,
\emph{Optimal exercise boundary for an American put option},
Applied Mathematical Finance
\textbf{5}
(1998),
107--116.


\bibitem{Kw}
Y.~K~Kwok,
\emph{Mathematical Models of Financial Derivatives},
Springer-Verlag (1998).


\bibitem{KW} 
Y.~K.~Kwok and L.~Wu, 
\emph{A Front-Fixing Finite Difference Method for the Valuation of American Options},
The Journal of Financial Engineering 
\textbf{6}
(1997),
83--97.


\bibitem{Lau} 
M.~Lauko, 
\emph{Numerical and analytical approximations of the early exercise boundary for the American put option},
Thesis, Comenius University, Bratislava 2009.




\bibitem{Mac}
L.~W.~MacMillan, 
\emph{Analytic approximation for the American put option},
Adv. in Futures Options Res.
\textbf{1}
(1986),
119--134.



\bibitem{MA}
R.~Mallier and G.~Alobaidi,
\emph{The American put option close to expiry},
Acta Mathematica Univ. Comenianae
\textbf{73}
(2004),
161--174.


\bibitem{M}
R.~Mynemi, 
\emph{The pricing of the American option},
Annal. Appl. Probab.
\textbf{2}
(1992),
1--23.



\bibitem{Se1}
D.~\v{S}ev\v{c}ovi\v{c},
\emph{Analysis of the free boundary for the pricing of an American call option},
Euro. Journal on Applied Mathematics
\textbf{12}
(2001),
25--37.

\bibitem{Se2}
D.~\v{S}ev\v{c}ovi\v{c},
\emph{An iterative algorithm for evaluating approximations to the optimal exercise boundary for a nonlinear Black-Scholes equation},
Canad. Appl. Math. Quarterly
\textbf{15}
(2007),
77--97.



\bibitem{SSC}
R.~Stamicar, D.~\v{S}ev\v{c}ovi\v{c} and J.~Chadam, 
\emph{The early exercise boundary for the American put near expiry: numerical approximation},
Canad. Appl. Math. Quarterly
\textbf{7} 
(1999),
427--444.


\bibitem{WDH}
P.~Wilmott, J.~Dewynne and S.~D.~Howison,
\emph{Option Pricing: Mathematical Models and Computation},
UK: Oxford Financial Press (1995).


\bibitem{Zhu2006} 
S.~P.~Zhu, 
\emph{A new analytical approximation formula for the optimal exercise boundary of American put options}, 
International Journal of Theoretical and Applied Finance
\textbf{9}(7)
(2006),
1141--1177.

\bibitem{Zhu2007}
S.~P.~Zhu and Z.~W.~He,
\emph{Calculating the early exercise boundary of American put options with an approximation formula},
International Journal of Theoretical and Applied Finance
\textbf{10}(7)
(2007),
1203--1227.

\bibitem{Zhu2008} 
S.~P.~Zhu, 
\emph{A Simple Approximation Formula For Calculating the Optimal Exercise 
Boundary of American Puts}, Preprint (2008).



\end{thebibliography}
\end{document}